\documentstyle[12pt]{article}
\begin{document}
\hspace{0.5cm}

\begin{center}
{\Large {\bf{ Geometric phases for non-degenerate and degenerate mixed states}\\
}}
\vspace*{1 cm}
\normalsize{
K. Singh$^0$ \footnotetext{Corresponding author. Email address: sciks@nus.edu.sg}, D. M. Tong, K. Basu, J. L. Chen and J. F. Du \\ 
\sl{ Department of Physics, National University of Singapore, 10 Kent Ridge Crescent, Singapore 119260}\\ 
\vspace*{1 cm}
}\end{center}
\abstract{This paper focuses on the geometric phase of general mixed states under unitary evolution. Here we analyze both non-degenerate as well as degenerate states. Starting with the non-degenerate case, we show that the usual procedure of subtracting the dynamical phase from the total phase to yield the geometric phase for pure states, does not hold for mixed states. To this end, we furnish an expression for the geometric phase that is gauge invariant. The parallelity conditions are shown to be easily derivable from this expression. We also extend our formalism to states that exhibit degeneracies. Here with the holonomy taking on a non-abelian character, we provide an expression for the geometric phase that is manifestly gauge invariant. As in the case of the non-degenerate case, the form also displays the parallelity conditions clearly. Finally, we furnish explicit examples of the geometric phases for both the non-degenerate as well as degenerate mixed states.}\\
\vspace{2cm}
\vspace{3cm}
\eject
\section{Introduction}
The notion of geometric phases was first addressed by Pancharatnam \cite{Pancharatnam} in his pioneering effort to compare the phases of two beams of polarized light. Although, his treatment was essentially classical, the notion of geometric phases was later shown to have important consequences for quantum systems. Indeed, Berry \cite{Berry}, demonstrated that quantum-mechanical systems could also acquire phases that are geometric in nature. He showed that, besides the usual dynamical phase, an additional phase that was related to the geometry of the state space was generated during an adiabatic evolution. Soon after, this phase was shown to be precisely the holonomy in a line bundle, establishing an even stronger link to the geometry of the state space. Essentially, Simon \cite{Simon} furnished an elegant geometrical interpretation   
of Berry's phase in which he considered a line bundle $L$ over the space of parameters $M$ of the system. More precisely, it was shown that $L$, if endowed with a particular connection, can reproduce the Berry's phase when closed loops in $M$ are lifted with respect to this connection. 
\par
It was subsequently shown that the adiabatic condition could be relaxed by considering the holonomy of a $U(1)$ bundle over the projective Hilbert space of the underlying system. By appealing to the notion of gauge invariance, Aharanov and Anandan \cite{Aharonov} showed that one could define a phase that was the same for all gauge-equivalent paths. Gauge-equivalent paths, here refer to the infinite number of possible evolutions (curves in the Hilbert space of the system) that project to a given closed curve in the projective Hilbert space of the system. 
\par
The abelian nature of the holonomy was also shown to be extendible to the non-abelian ones. Wilczek and Zee \cite {Wilczek} 
 showed that for Hamiltonians that exhibited degeneracies, the adiabatic evolution admitted holonomies that were generally non-abelian. The non-adiabatic generalization of this was subsequently furnished by Anandan \cite{Anandan}.  
\par
It is interesting to note that the notion of geometric phases has taken on an important role in the area of quantum  computation \cite{Ekert1}. Indeed, it has been demonstrated that one can implement quantum logic gates based solely on the concept of geometric phase. The geometrical nature of the phase makes such a realization fault-tolerant and thus more robust towards noise. The application has also been proposed within the context of non-abelian holonomies. Termed Holonomic Quantum Computation, the program provides an all-geometrical approach to quantum information processing \cite{Zanardi}.       
\par
While many of the propositions in this area has been centered around pure states, the need to address the issue of geometric     phase for mixed states is rapidly gaining prominence. Systems that are interacting with other systems, the environment for instance,  are generally described by mixed states. Again, the strong focus in this area is fueled by the promise of realizing quantum logic gates under more realistic physical conditions. 
\par
Now, among the first to introduce the notion of geometric phase for mixed states was Uhlmann \cite{Uhlmann}. By considering a purification  and the notion of parallelity, he furnished a definition for the geometric phase for mixed quantal states. Recently Sj\"oqvist et.al. \cite{Sjoqvist} have introduced a new formalism that defines the geometric phase in the context of quantum interferometry. Using the standard setup of the Mach-Zender Interferometer, they provided a procedure for  calculating the geometric phase of mixed state undergoing unitary evolutions. They have also furnished the necessary and sufficient conditions for parallel transporting mixed states under such evolutions. 
\par              
Recently, it was pointed out in Ref.\cite{Slater} that the two approaches, in general, are not equivalent . More recently, Ericcson et.al. \cite{Ericsson} have shown that the conditions of parallelity used in the two approaches lead to generically distinct phase holonomy effects for entangled systems undergoing certain local unitary transformations.        
\par
In this paper, we consider the geometric phase of general mixed states under unitary evolution. Here we consider both non-degenerate as well as degenerate states. Starting with the non-degenerate case, we show that the removal of the dynamical phase from the total phase is not a trivial subtraction as is the case for pure states. In particular, we show that such a procedure is generally not gauge invariant. We then proceed to furnish an expression that is manifestly gauge invariant. The parallelity conditions are shown to be easily derivable from this expression. We also extend our formalism to states that exhibit degeneracies. Parallelity conditions for such states are also obtained. Finally, we furnish explicit examples of the geometric phases for both the non-degenerate as well as degenerate mixed states.       

\section{The Non-adiabatic Geometric Phase}
In this section we begin by briefly reviewing some basic facts about the non-adiabatic geometric phase. Here, we will concentrate on the abelian case, starting with pure states (\cite{Aharonov},\cite{Mukunda}, \cite{Bohm}).   
We begin by considering a quantum system $S$ with states belonging to the Hilbert space $\cal H$. The space of physical states corresponds to the non-zero vectors in $\cal H$ that are rendered equivalent under the relation
\begin{eqnarray}
\vert \psi_1> \sim \vert\psi_2>  \quad {\rm iff}\quad  \vert \psi_2> = e^{i\theta}\vert \psi_1>.
\label{ray}
\end{eqnarray}
where $\theta$ is any real number. Technically, for an $n$-state system with dim($\cal H$) =$n$ this space is simply a complex projective space of dimension $(n-1)$: 
 \begin{eqnarray}
{\cal P}({\cal H}) = \frac{{\cal H} -\{{ 0} \}}{{\cal C} -\{ 0 \}}\equiv CP^{n-1}
\label{CPn}
\end{eqnarray}

In the language of fiber bundles, the construct of interest is a principal bundle over ${\cal P}({\cal H})$ with a fiber that is isomorphic to $U(1)$. Here the bundle space $P$ corresponds to the space of normalized state vectors ${\cal H} -\{{ 0} \}$ with a natural projection 
\begin{eqnarray}
\pi : P \to {\cal P}({\cal H})\equiv {\cal M} \quad {\rm realized \ through} \quad \pi: \vert\psi> \to \vert\psi><\psi\vert.
\label{pi}
\end{eqnarray} 
Now the unitary evolution of the state vector 
\begin{eqnarray}
\vert\psi(0)> \to \vert\psi(t)> = {\cal U} (t)\vert\psi(0)>  
\label{evolution}
\end{eqnarray}
produces a path $C:[0,\tau]\to P$ in the bundle space. This is dictated by the Hamiltonian via the Schr\"odinger equation.  
The corresponding path $C_0$ in the space of physical states is obtained by projecting the path $C$ through the map $\pi$, {\it i.e.} $\pi(C)=C_0$. Moreover, we say that the evolution is cyclic if the path in $\cal M$ is closed; {\it i.e.} if 
$\vert\psi(\tau)><\psi(\tau)\vert= \vert\psi(0)><\psi(0)\vert$. It is important to note that there are an infinite number of paths in $P$ that project to the same path in $\cal M$. Indeed, if $C$ and $C'$ are given by $\vert \psi(t)>$ and   $\vert \psi'(t)>= e^{if(t)}\vert \psi(t)>$ respectively for any arbitrary real function $f(t)$, then it is easy to see that they define the same path $C_0$ in $\cal M$ under the projection $\pi$. For a cyclic evolution, the total phase acquired corresponds to an element of the fiber {\it i.e.} $e^{i\phi_T}\in U(1)$. The total phase  $\phi_T$, here, being the argument of $<\psi(0)\vert \psi(\tau)>$:
 \begin{eqnarray}
{\rm total \ phase} \ \phi_T[C] 
& = & {\rm arg}\{ <\psi(0)\vert \psi(\tau)>\}\nonumber \\
& = & {\rm arg}\{ <\psi(0)\vert {\cal U} (t)\vert\psi(0)>\}
\label{total-phase}
\end{eqnarray}
In general, this depends on the path $C$ taken in the bundle space $P$, which in turn, is defined by ${\cal U} (t)$. This is reminiscent of the Pancharatnam analysis which compares the relative phase between two states. Aharonov and Anandan \cite{Aharonov} showed that if a quantity (called the dynamical phase), 
\begin{eqnarray}
\phi_D[C] & = & -\int_{0}^{\tau}<\psi(t)\vert H\vert \psi(t)> dt \nonumber \\
& = & -i\int_{0}^{\tau}<\psi(0)\vert {\cal U} (t)^{\dag} \dot {\cal U} (t)\vert \psi(0)> dt 
\label{dynamical-phase}
\end{eqnarray}
is subtracted from the total phase (\ref{total-phase}) then the resulting term 
\begin{eqnarray}
\phi_G  = {\rm arg}\{ <\psi(0)\vert {\cal U} (t)\vert\psi(0)>\} +  i\int_{0}^{\tau}<\psi(0)\vert {\cal U} (t)^{\dag} \dot {\cal U} (t)\vert \psi(0)> dt 
\label{geometric-phase}
\end{eqnarray}
turns out to be a functional of $C_0$ only. This is the notion of gauge invariance as emphasized in Ref \cite{Mukunda}. Since the path $C$ in $P$ from a given intial state $\vert \psi(0)>$ is solely determined by ${\cal U}(t)$ we can label the path by the latter. Here the statement of gauge invariance for the geometric phase for all paths differing by a $U(1)$ phase reads as::
\begin{eqnarray}
\phi_G[{\cal U} (t)e^{if(t)}]= \phi_G[{\cal U} (t)]
\label{gauge-inv}
\end{eqnarray}
for any arbitrary real function $f(t)$. Thus equation (\ref{geometric-phase}) furnishes a gauge-invariant expression for the geometric phase of a pure state. It is worth noting that arbitrariness of $f(t)$ allows one to impose a condition known as the parallel transport condition:
\begin{eqnarray}
<\psi(0)\vert {\cal U} (t)^{\dag}\dot {\cal U} (t)\vert \psi(0)>=0
\label{parallel-transport}
\end{eqnarray}
which renders the dynamical phase zero. Under this condition the geometric phase is just the total phase.
\par
For mixed states, Sj\"oqvist et. al. \cite{Sjoqvist} showed that the total phase acquired by the system under unitary evolution is a natural generalization of the pure state case. Indeed, by considering a Mach-Zender interferometer, in which the interfering particles carry an additional degree of freedom (for which the state is mixed), they showed that under a unitary evolution
\begin{eqnarray}
\rho(0)\to \rho(\tau) = {\cal U} (\tau)\rho(0){\cal U} (\tau)^{\dag}
\label{u-evolute}
\end{eqnarray}
the interference pattern of the intensity profile takes the form of 
\begin{eqnarray}
I \propto 1 + \vert {\rm Tr}({\cal U} (\tau)\rho(0))\vert \cos [\chi - {\rm arg}\{ {\rm Tr}({\cal U} (\tau)\rho(0))\} ].
\label{interference}
\end{eqnarray}
Here $\chi$ is a variable relative $U(1)$ phase in one of the interferometer beams. From the shift in the interferometric pattern, they surmised that the total phase acquired by a mixed state under unitary evolution is
 \begin{eqnarray}
\gamma_T  & = &{\rm arg}\{ {\rm Tr}({\cal U} (\tau)\rho(0))\}
\label{total-mix}.
\end{eqnarray}
In a basis in which the density matrix $\rho (0)$ is expreseed as,
\begin{eqnarray}
\rho(0)=\sum_{k=1}^{N}\omega_k\vert k><k\vert, \quad {\rm{dim}}({\cal H})=N  
\label{density}
\end{eqnarray}
the total phase is just the weighted sum of the pure state phases: 
\begin{eqnarray}
\gamma_T  & = &{\rm arg}\{ {\rm Tr}({\cal U} (\tau)\rho(0))\}\nonumber\\
& = & {\rm arg}\{\sum_{k=1}^{N}\omega_k<k\vert {\cal U} (\tau)\vert k> \}.
\label{total-mix}
\end{eqnarray}
In Ref. \cite{Sjoqvist}, the parallel transport condition for non-degenerate state was imposed by requiring  that ${\rm Tr}[\rho(t){\cal U}(t+dt){\cal U} (t)^{\dag}]$ be real and positive
which in turn leads to
\begin{eqnarray}
{\rm Tr}[\rho(t)\dot {\cal U} (t){\cal U} (t)^{\dag}] \equiv {\rm Tr}[\rho(0){\cal U} (t)^{\dag }\dot {\cal U} (t)] = 0
\label{PT-mixed}
\end{eqnarray}
or equivalently 
\begin{eqnarray}
\sum_{k=1}^{N}\omega_k <k\vert {\cal U} (t)^{\dag}\dot {\cal U} (t)\vert k>=0.
\label{PT-mixed2}
\end{eqnarray}
Under this condition the dynamical phase 
\begin{eqnarray}
\gamma_D & = & - \int_{0}^{\tau} dt {\rm Tr}(\rho(t)H(t))\nonumber \\
& = & - i \int_{0}^{\tau} dt {\rm Tr}(\rho(0){\cal U}(t)^{\dag}\dot{{\cal U}}(t))
\label{dy-phase}
\end{eqnarray}
vanishes identically. It is important to note that if condition (\ref{PT-mixed2}) is not met then the dynamical phase term in eqn.(\ref{dy-phase}) cannot be removed from the total phase term, (\ref{total-mix}) by a simple subtraction; since the latter is defined by a weighted sum of phase factors while the former is a weighted sum of phases.  
The authors of Ref. \cite{Sjoqvist} have also asserted that conditions (\ref{PT-mixed}) or (\ref{PT-mixed2}) while necessary,  is not sufficient. Instead  they proposed stronger conditions 
\begin{eqnarray}
 <k\vert {\cal U} (t)^{\dag}\dot {\cal U} (t)\vert k>=0 \qquad \forall k= 1,2,...N
\label{PT-mixed3}
\end{eqnarray}
in which all the constituent pure states in the mixture are required to be parallel transported independently. 
\par
In the following section we elucidate the nature of these conditions. In particular, we furnish these by first constructing an expression for the geometric phase that is manifestly gauge invariant. This is first done for the non-degenerate case in which we show that the removal of the dynamical phase requires the use of the stronger conditions noted above. 
We also extend our analysis to states that are degenerate. Here, we also provide a generalized version of the geometric phase that is applicable when the state is multiply degenerate. The corresponding parallel transport conditions are also furnished   

\section{Gauge-Invariance and Holonomy in Mixed States}
\par
To address the issue of gauge invariance in the context of holonomy for mixed states, it is instructive to consider the approach of Boya et.al.{\cite{Boya}}. Here one considers the little group of a particular mixed state density matrix which can be taken as the gauge group for the system that is undergoing unitary evolution. It should be noted that the space of states accessible through unitary evolutions from a reference state does not cover the entire state space. For instance, if we are considering a spin-$\frac{1}{2}$ system, the state space consists of all points on and inside the Bloch sphere.  However, if we start with a particular mixed state and assume that it is undergoing unitary evolution then the accessible states lie on a spherical shell with radius equal to the magnitude of the initial Bloch vector. In determining the relevant bundle structure, we appeal to an important theorem for spaces with transitive group action. Essentially, it is well known that if a group $G$ acts transitively on a space ${\cal M}$ then for each $x\in {\cal M}$ there is a bijection between $G/G_x$ and ${\cal M}$ where $G_x$ denotes the little group or the stability group at the point $x$ \cite{Thomas}. For the problem at hand, we are interested in the little group associated with the density matrix at time $t=0$. The theorem basically assures us that the  orbit space under unitary evolution is isomorphic to the coset space, $G/G_x$. More importantly, if the little group is closed then ($G$, $G_x$, $\pi$, $G/G_x$), where $\pi: G \to G/G_x$ is the canonical map which sends each element of the group to the cosets, serve to define a bundle structure. Here $G$ is taken as the bundle space, $G/G_x$ the base space, $G_x$ the fiber and $\pi$ the projection. 
\par 
To carry the analysis further, it is instructive to consider a density matrix in its diagonal basis (\ref{density}). 
Let us start with the case in which the $N$-state  density matrix has eigenvalues that are distinct, {\it i.e.} non-degenerate. Then the little group in this case is simply 
\begin{eqnarray}
G_x = \underbrace{U(1)\times U(1)\cdots \times U(1)}_{N} \equiv \sum_{n=1}^{N}e^{i\theta_n}\vert n><n\vert
\label{little-gp}
\end{eqnarray}
where $\{\theta_n \}_{n=1,2,...N}$ are arbitrary parameters or phases. The group $G$ in this context is $U(N)$. It is easy to see that under 
\begin{eqnarray}
{\cal U}(t)\in U(N) \to {\cal U}'(t)={\cal U}(t)\sum_{n=1}^{N}e^{i\theta_n(t)}\vert n><n\vert
\label{gauge}
\end{eqnarray}
the orbit of the density matrix remains unchanged: 
\begin{eqnarray}
\rho(0)\to \rho'(t) = {\cal U}'(t)\rho(0){\cal U}'(t)^{\dag} = {\cal U} (t)\rho(0){\cal U} (t)^{\dag}=\rho(t).
\label{rho-evolute}
\end{eqnarray}
In other words, there are infinite number of orbits in $G$ that correspond to  same path for $\rho (t)$. Now, unlike the pure state case (see (\ref{geometric-phase})), the removal of the dynamical phase from the total phase does not render the resulting functional gauge invariant. Indeed under a gauge transformation (\ref{gauge}) the total phase (\ref{total-mix}) transforms as 
 \begin{eqnarray}
\gamma_T \to \gamma_T' & = & {\rm arg} \{{\rm Tr}[\rho(0){\cal U}'(\tau)]\}\nonumber \\
& = & {\rm arg}\{\sum_{k}\omega_k<k\vert {\cal U}(\tau)\vert k>e^{i\theta_k (t)}\}\label{gammaT}
\end{eqnarray}
while the dynamical phase (\ref{dy-phase}) transforms as
\begin{eqnarray}
\gamma_D \to \gamma_D' & = & - i \int_{0}^{\tau}  dt \ {\rm Tr}(\rho(0){\cal U}'(t)^{\dag}\dot{{\cal U}'}(t))\nonumber\\
&=& - i \int_{0}^{\tau} dt \ {\rm Tr}(\rho(0){\cal U}(t)^{\dag}\dot{{\cal U}}(t)) + \sum_k\omega_k\theta_{k}(\tau).
\label{gammaD}
\end{eqnarray}
It is evident from (\ref{gammaT}) and (\ref{gammaD}) that the $\theta$-dependence in total phase cannot be removed by simply
subtracting the dynamical phase term. It is also easy to see that only in the pure-state case, with only one non-zero $\omega$, do we have gauge-invariance. 
\par
To alleviate this problem we propose a functional, of the following form:
\begin{eqnarray}
\gamma_G [{\cal U}] = {\rm arg}\{\sum_k [\omega_k <k\vert {\cal U}(\tau)\vert k> e^{-\int_{0}^{\tau}dt <k \vert{\cal U}(t)^{\dag}\dot {\cal U}(t)\vert k>}]\}
\label{gauge-in-phase}
\end{eqnarray}
which is manifestly gauge invariant (as in (\ref{geometric-phase})). Indeed, the $\theta$ term generated in (\ref {gammaT})
under a gauge transformation (\ref {gauge}) is exactly cancelled by a term arising from the exponential term in (\ref {gauge-in-phase}). It is easy to check that the above expression corresponds to the geometric phase in the case when the state is pure. Gauge invariance also assures us that the expression depends only on the path in the state space. This expression, thus furnishes us with a gauge-invariant definition for the geometric phase of  mixed states that are non-degenerate. It is worth noting that when the stronger conditions (\ref{PT-mixed3}) are imposed the resulting expression is nothing but the original total phase term, which means, they are justifiably parallel transport conditions; since only under parallel transport is the geometric phase the same as the total phase. 
\par      
The above analysis holds only for mixed states that are non-degenerate. For degenerate mixed states, it is well known that the holonomy group can take on a non-abelian character \cite{Boya}. To this end, let us consider a density matrix that has the following diagonal form:
\begin{eqnarray}
\rho(0)=\sum_{k=1}^{N} \omega_k\vert k><k\vert 
\label{matrix}
\end{eqnarray}
in which $n$ of the eigenvalues are degenerate {\it i.e.} $\omega_1 = \omega_2 = \cdots =\omega_{n}$ with the rest distinct. Here the Hilbert space $\cal H$ can be regarded as the direct sum of two subspaces, ${\cal H}_n \oplus {\cal H}_m$ of dimensions $n$ and $m=N-n$.  
 The little group of ${\rho}(0)$ is  non-abelian and has the following form:
\begin{eqnarray}
G_x =U(n)\times\underbrace{U(1)\times U(1)\times\cdots \times U(1)}_{m}
\label{mat}
\end{eqnarray}
or in the $\{\vert k>\}_{k=1,2,...N}$ basis it takes the form
\begin{eqnarray}
{\cal V}(t)=\left\lgroup\begin{array}{cccccc}
\left\lgroup\matrix{
~ &~ &~\\
~ &\alpha (t) &~\\
~ &~ &~\\
}\right\rgroup
\matrix{
~ &~ &~ &~ &~ &~ &~\\
~ &~ &~  &~ &~ &~ &~\\
~ &~ &~  &~ &~ &~ &~\\
}\\
\matrix{
~ &~ &~ &~ &~ \\
~ &~ &~ &~ &~\\
~ &~ &~ &~ &~\\
}
\left\lgroup\matrix{
e^{i\beta_{n+1}(t)} &~ &~\\
~ &\ddots &~\\
~ &~ &e^{i\beta_{N}(t)}\\
}\right\rgroup
\end{array}\right\rgroup.
\end{eqnarray}
Here the only non zero terms are the elements of $\alpha$, an $n\times n$ unitary submatrix and exponential of the $\beta$'s along the  diagonal $(m\times m)$-submatrix. It is easy to see that under a gauge transformation
\begin{eqnarray}
{\cal U}(t) \to {\cal U}'(t)= {\cal U}(t){\cal V}(t)
\label{GT}
\end{eqnarray}
the path taken by the density matrix remains unchanged. It is worth noting that, since both ${\cal U}(t)$ and ${\cal U}'(t)$ are the time evolution unitary operators for the system, we must require that ${\cal U}(0) = {\cal U}'(0)= 1$. As a consequence, the element of the little group ${\cal V}(t)$ must also satisfy ${\cal V}(0)=1$.  Now, consider the total phase term (\ref{total-mix}) under such a transformation: 
\begin{eqnarray}
\gamma_T \to \gamma_T' & = & {\rm arg} \{{\rm Tr}_{\cal H}[\rho(0){\cal U}'(\tau)]\}\nonumber \\
& = & {\rm arg}\{\sum_{k}\omega_k<k\vert {\cal U}(\tau){\cal V}(\tau)\vert k>.
\label{GT1}
\end{eqnarray}
Following the non-denegerate case, we note that in constructing a gauge-invariant functional an expression of the form 
 \begin{eqnarray}
\gamma[{\cal U}] = {\arg} \{{\rm Tr}_{\cal H}(\rho(0){\cal U}(\tau){\cal F} [{\cal U};\tau])\} \label{GT2}
\end{eqnarray}
with a functional ${\cal F}[{\cal U};\tau]$ that transforms as 
 \begin{eqnarray}
{\cal F}[{\cal U};\tau] \to {\cal V}(\tau)^{\dag} {\cal F}[{\cal U};\tau]
 \label{GT3}
\end{eqnarray}
under a gauge transformation would be required. To identify this functional, we first observe that in the non-degenerate case, the corresponding functional $e^{-\int_{0}^{\tau}dt <k \vert{\cal U}(t)^{\dag}\dot {\cal U}(t)\vert k>}$ can be obtained by fixing the $\theta$ terms in a general evolution matrix (\ref {gauge}) through the conditions (\ref{PT-mixed3}) for $\cal U'$. This yields
 \begin{eqnarray}
\theta_k(t) = i\int_{0}^{t}dt' <k \vert{\cal U}(t')^{\dag}\dot {\cal U}(t')\vert k> \qquad k=1,2,...,N
\label{GT4}
\end{eqnarray}
which upon substitution into (\ref{gauge}) leads to the required form. In the degenerate case, we have $n^2+m$ arbitrary parameters that need to be fixed; $n^2$ for the $\alpha$ matrix and $m$ parameters  corresponding to the $\beta$'s. We can do this by imposing
\begin{eqnarray}
\begin{array}{ccccccc}
<\mu\vert {\cal U}'^{\dag}\dot{{\cal U}'}\vert \nu> & = & 0 \quad \mu,\nu &= & 1,2,...n  \label{GT5}\\
<k\vert {\cal U}'^{\dag}\dot{{\cal U}'}\vert k> & = & 0 \quad k & = & n+1,n+2,...N
\label{GT6}
\end{array}
\end{eqnarray}
where we have defined basis vectors $\{\vert \mu>\}_{\mu=1,2,...n}$ and $\{\vert k>\}_{k=n+1,n+2,...N}$ to designate subspaces 
${\cal H}_n$ and ${\cal H}_m$ respectively. With 
\begin{eqnarray}
{\cal U}'={\cal U}(t) \left\lgroup\begin{array}{cccccc}
\left\lgroup\matrix{
~ &~ &~\\
~ &{\alpha}(t) &~\\
~ &~ &~\\
}\right\rgroup
\matrix{
~ &~ &~ &~ &~ &~ &~\\
~ &~ &~  &~ &~ &~ &~\\
~ &~ &~  &~ &~ &~ &~\\
}\\
\matrix{
~ &~ &~ &~ &~ \\
~ &~ &~ &~ &~\\
~ &~ &~ &~ &~\\
}
\left\lgroup\matrix{
e^{i\beta_{n+1}(t)} &~ &~\\
~ &\ddots &~\\
~ &~ &e^{i\beta_{N}(t)}\\
}\right\rgroup
\end{array}\right\rgroup.
\end{eqnarray}
conditions (\ref{GT5}) lead to the following equations:
\begin{eqnarray}
\dot{\alpha}_{\mu\nu}(t)& =& -({\cal U}(t)^{\dag}\dot{{\cal U}}(t))_{\mu\lambda} \alpha_{\lambda \nu}(t)\label{GT5a}\\
\dot{\beta}_k(t) & = & i ({\cal U}(t)^{\dag}\dot{{\cal U}}(t))_{kk} \label{GT5b}
\end{eqnarray}
where  $\alpha _{\mu\nu}$ are the elements of the $\alpha$ matrix. Equations (\ref{GT5a}) and (\ref{GT5b}) can be formally integrated to yield
\begin{eqnarray}
\alpha_{\mu\nu}[{\cal U};t] &=& <\mu\vert {\rm P}\ {\rm {exp}}- \int_{0}^{t}dt' {\cal U}(t')^{\dag}\dot{\cal U}(t')\vert\nu> \quad \mu,\nu = 1,2,...n\label{GT7}\\
\beta_k [{\cal U};t]& = & i\int_{0}^{t} dt' <k\vert{\cal U}(t')^{\dag}\dot{\cal U}(t') \vert k> \quad k = n+1,n+2,...N \label{GT8} 
\end{eqnarray}
where P denotes path ordering. We can now choose 
\begin{eqnarray}
{\cal F}[{\cal U};\tau]=\left\lgroup\begin{array}{cccccc}
\left\lgroup\matrix{
~ &~ &~\\
~ &\alpha [{\cal U};\tau] &~\\
~ &~ &~\\
}\right\rgroup
\matrix{
~ &~ &~ &~ &~ &~ &~ &~ &~\\
~ &~ &~  &~ &~ &~ &~ &~ &~\\
~ &~ &~  &~ &~ &~ &~ &~ &~\\
}\\
\matrix{
~ &~ &~ &~ &~ \\
~ &~ &~ &~ &~\\
~ &~ &~ &~ &~\\
}
\left\lgroup\matrix{
e^{i\beta_{n+1}[{\cal U};\tau]} &~ &~\\
~ &\ddots &~\\
~ &~ & e^{i\beta_{N}[{\cal U};\tau]}\\
}\right\rgroup
\end{array}\right\rgroup.
\end{eqnarray}
It is worth noting that ${\cal F}[{\cal U};\tau]$ is block diagonal in the chosen basis and that it can be written as direct sum form:
 \begin{eqnarray}
{\cal F}[{\cal U};\tau] = {\cal F}_{{\cal H}_n}[{\cal U};\tau] \oplus {\cal F}_{{\cal H}_m}[{\cal U};\tau] \label{GT9}
\end{eqnarray}
with ${\cal F}_{{\cal H}_n}[{\cal U};\tau]=\alpha [{\cal U};\tau]$ and ${\cal F}_{{\cal H}_m}[{\cal U};\tau]= {\rm diag}(e^{i\beta_{n+1}[{\cal U};\tau]},e^{i\beta_{n+2}[{\cal U};\tau]},...,e^{i\beta_{N}[{\cal U};\tau]})$.
\par
 We next demonstrate that expression (30) is gauge invariant. Indeed, under 
\begin{eqnarray}
{\cal U}\to {\cal U}'={\cal U}(t) \left\lgroup\begin{array}{cccccc}
\left\lgroup\matrix{
~ &~ &~\\
~ &{\cal V}_1(t) &~\\
~ &~ &~\\
}\right\rgroup
\matrix{
~ &~ &~ &~ &~ &~ &~\\
~ &~ &~  &~ &~ &~ &~\\
~ &~ &~  &~ &~ &~ &~\\
}\\
\matrix{
~ &~ &~ &~ &~ \\
~ &~ &~ &~ &~\\
~ &~ &~ &~ &~\\
}
\left\lgroup\matrix{
e^{i\theta_{n+1}(t)} &~ &~\\
~ &\ddots &~\\
~ &~ &e^{i\theta_N(t)}\\
}\right\rgroup
\end{array}\right\rgroup.
\end{eqnarray}
where ${\cal V}_1(t)$ is an arbitrary $n\times n$ unitary matrix, we note that 
\begin{eqnarray}
{\cal F}_{{\cal H}_n}[{\cal U};\tau] & \to & {\cal V}_1({\tau})^{\dag}{\cal F}_{{\cal H}_n}[{\cal U};\tau]\label{GT11}\\
{\cal F}_{{\cal H}_m}[{\cal U};\tau] & \to & {\rm diag}(e^{i\theta_{n+1}(\tau)},e^{i\theta_{n+2}(\tau)},...,e^{i\theta_N(\tau)})^{\dag}{\cal F}_{{\cal H}_m}[{\cal U};\tau]
 \label{GT20}
\end{eqnarray}
which implies that $\gamma [{\cal U}]\to \gamma [{\cal U}'] = \gamma[(\cal U)]$ (see Appendix for details). Thus expression 
(\ref{GT2}) furnishes a natural generalization of the geometric phase for a degenerate mixed state. It is easy to see that the conditions (\ref{GT5}) here serve as the parallelity conditions.
\par
It is interesting to note that the above analysis can be extended to density matrices that are multiply degenerate. For instance if ${\rho}(0)$ has eigenspaces ${\cal H}_1$, ${\cal H}_2$ ...${\cal H}_m$ with degeneracies $n_1$, $n_2$ ,...$n_m$ respectively, then the Hilbert space of the system can be written as a direct sum of degenerate subspaces:
\begin{eqnarray}
{\cal H}_N = {\cal H}_1 \oplus {\cal H}_2\oplus ... \oplus {\cal H}_m \quad {\rm{ with}} \quad N =\sum_{k=1}^{m}n_k .
 \label{GT21}
\end{eqnarray}   
The corresponding ${\cal F}$-functional for ${\cal H}_{N}$ can also be written as the direct sum of ${\cal F}$-functionals define on the subspaces:
\begin{eqnarray}
{\cal F}_{{\cal H}_N}[{\cal U,\tau}] = {\cal F}_{{\cal H}_1}[{\cal U,\tau}] \oplus {\cal F}_{{\cal H}_2}[{\cal U,\tau}],\oplus, ..., \oplus {\cal F}_{{\cal H}_m}[{\cal U,\tau}]
 \label{GT22}
\end{eqnarray}
where the components of ${\cal F}_{{\cal H}_k}[{\cal U,\tau}]$
are given by
\begin{eqnarray}
{\cal F}_{{\cal H}_k} [{\cal U,\tau}]_{{\mu}_k{\nu}_k}= <\mu_{k}\vert{\rm {P}} \ {\rm{exp}}- \int_{0}^{\tau} \ dt \ {\cal U}(t)^{\dag} \dot {\cal U}(t) \vert \nu_{k}> \quad \mu_k,\nu_k = 1,2,...,{\rm{dim}}({\cal H}_k)
\label{GT23}
\end{eqnarray}
if dim $({\cal H}_k) \neq 1$ and 
\begin{eqnarray}
{\cal F}_{{\cal H}_k} [{\cal U,\tau}]= e^{- \int^{\tau}_{0} \ dt \ <\mu_k\vert {\cal U}(t)^{\dag}\dot {\cal U}(t) \vert \mu_k>} 
\label{GT24}
\end{eqnarray}
otherwise. The geometric phase is then calculated directly from
\begin{eqnarray}
\gamma[{\cal U}] = {\rm {arg}}\{{\rm {Tr}}_{{\cal H}_N}({\rho}(0){\cal U}(\tau){\cal F}_{{\cal H}_N}[{\cal U};\tau])\}
\label{GT25}
\end{eqnarray}
\section{Explicit Examples}
In this section we furnish explicit examples of the geometric phase for both the non-degenerate as well as the degenerate mixed states. We begin with the non-degenerate case. To this end, we consider the simplest case of a spin- 
$\frac{1}{2}$ system with a density matrix given by
\begin{eqnarray}
\rho = \frac{1}{2}({\bf 1} + {\bf r}\cdot {\bf \sigma} )
\label{r012}
\end{eqnarray}
where $\bf r$ is a three-dimensional Bloch vector and $\bf \sigma$ the Pauli matrices. For the purpose of computation, we set the Bloch vector of the initial state to be $(r\sin\theta,0, r\cos\theta)$, ${{i.e.}}$   
\begin{eqnarray}
\rho(0)=\frac{1}{2} \left(\begin{array}{cc}
1+r\cos\theta & r\sin\theta \\
r\sin\theta & 1-r\cos\theta
\end{array}\right).
\label{r0m}
\end{eqnarray}
If we subject the system to a unitary evolution 
\begin{eqnarray}
\rho(0) \to \rho(t) = {\cal U}(t)\rho(0){\cal U}(t)^{\dag}\quad {\rm with} \quad {\cal U}(t)= \exp({-\frac{it}{2}\sigma_3}) 
\label{dmat}
\end{eqnarray}  
the Bloch vector precesses about the $z$-axis (at a constant polar angle of $\theta$). For cyclic evolution, the parameter $t$ takes values $[0,2\pi]$. Now, the matrix (46) can be diagonalized: 
\begin{eqnarray}
\rho(0)=\frac{1}{2} \left(\begin{array}{cc}
1+r & 0 \\
0 & 1-r
\end{array}\right)
\label{r1}
\end{eqnarray}  
with respect to its eigenvectors:
\begin{eqnarray}
\vert 1 > = \left(\begin{array}{c}
\cos\theta \\
\sin\theta
\end{array}\right) \qquad \vert 2 > = \left(\begin{array}{c}
\sin\theta \\
-\cos\theta
\end{array}\right).
\label{r2}
\end{eqnarray}  
In (\ref{r1}) we assume $0<r<1$. The geometric phase can obtained by evaluating (\ref{gauge-in-phase}):
\begin{eqnarray} 
\gamma_G [{\cal U}] & = & {\rm arg}\left[-\left(\frac{1+r}{2}\right) e^{i\pi\cos\theta} -\left(\frac{1-r}{2}\right) e^{-i\pi\cos\theta} \right] \\
& = &- \arctan (r \tan \frac{\Omega}{2}) 
\label{r3}
\end{eqnarray}  
where $\Omega = 2 \pi (1-\cos\theta)$ is the solid angle subtended by the Bloch vector. It is interesting to note that the same expression has been furnished in \cite{Sjoqvist} where parallel transport conditions were employed. Here we have obtained the geometric phase without the need for such conditions. 
\par   
We now turn to the degenerate case. Here we choose a three-state density matrix given by:
\begin{eqnarray}
\rho =\frac{1}{3} (\textbf{1}+ \sum_{i=1}^{8}{r}_i {\lambda}_i)  \label{r4}  
\end{eqnarray}
where $\{r_i\}$ are the components of an eight dimensional vector and $\{\lambda_i\}$  are the generators of $SU(3)$:
\begin{displaymath}
\lambda_{1} =\left[
	\begin{array}{ccc}
		0& 1 & 0  \\
		1 & 0 & 0\\
		0& 0 & 0  \\
		
	\end{array}\right] \quad  \lambda_{2} =\left[
	\begin{array}{ccc}
		0& -i & 0  \\
		i & 0 & 0\\
		0& 0 & 0  \\
		
	\end{array}\right] \quad  \lambda_{3} =\left[
	\begin{array}{ccc}
		1& 0 & 0  \\
		0 &-1 & 0\\
		0& 0 & 0  \\
		
	\end{array}\right] \quad   \lambda_{4} =\left[
	\begin{array}{ccc}
		0& 0 & 1  \\
		0 & 0 & 0\\
		1& 0 & 0  \\
		
	\end{array}\right] \\\end{displaymath}
\\\begin{displaymath}
\lambda_{5} =\left[
	\begin{array}{ccc}
		0&0 & -i  \\
		0 & 0 & 0\\
		i& 0 & 0  \\
		
	\end{array}\right] \quad  \lambda_{6} =\left[
	\begin{array}{ccc}
		0& 0 & 0  \\
		0 & 0 & 1\\
		0& 1 & 0  \\
		
	\end{array}\right] \quad \lambda_{7} =\left[
	\begin{array}{ccc}
		0& 0 & 0  \\
		0 &0 &-i\\
		0& i & 0  \\
		
	\end{array}\right]  \quad  \lambda_{8} =\frac{1}{\sqrt{3}}\left[
	\begin{array}{ccc}
		1& 0 & 0  \\
		0& 1 & 0\\
		0& 0 & -2  \\
	\end{array}\right].\\
\end{displaymath}\\
For definiteness we choose the initial state to be 
\begin{eqnarray}
\rho(0)=\left[
	\begin{array}{ccc}
		\omega & 0 & 0  \\
		0 & \omega & 0\\
		0& 0& 1-2\omega  \\
	\end{array}\right] \qquad {\rm with} \quad 0<\omega< \frac{1}{2} . 
\label{r5}
\end{eqnarray}\\
Here the degeneracy is manifest with the little group given by $U(2)\times U(1)$. It is interesting to note that in the above basis for the $SU(3)$ generators, the subgroup $SU(2)\times U(1)$ is spanned by $\{\lambda_1,\lambda_2,\lambda_3,\lambda_8\}$.  

For an explicit evaluation of the geometric phase we consider a unitary matrix of $SU(3)$ given by 
\begin{eqnarray}
{\cal U}(t) = e^{-it X}, \qquad{\rm where}\quad X = a\lambda_8 + b\lambda_4
\label{r6}
\end{eqnarray}
where $a$ and $b$ are arbitrary real parameters. In a matrix form, we have 
\begin{eqnarray}
{\cal{U}}(t)=e^{\frac{ia\tau}{2\sqrt{3}}} \left[
	\begin{array}{ccc}
		c\, \cos\frac{c t}{2}-i\sqrt{3}a\, \sin\frac{c t}{2} & 0 & \frac{-2bi}{a}\sin\frac{ct}{2}  \\
		0 & e^{\frac{-i\sqrt{3}a\, t}{2}} & 0\\
		 \frac{-2bi}{a}\sin\frac{c t}{2}& 0&  c\, \cos\frac{c t}{2}+i\sqrt{3}a\, \sin\frac{c t}{2} \\
	\end{array}\right] \label{r7} \end{eqnarray}
where $c = \sqrt{3a^2 + 4b^2}$. The cyclicity conditions can be obtained by restricting $t$ to lie within the interval $[0,\tau]$ where $\tau$ is evaluated by requiring that $r_i(\tau) = r_i(0)$ with
\begin{eqnarray}
r_i(t) &=& \frac{3}{2} {\rm {Tr}} [\rho(t)\lambda_i] \\
&=& \frac{3}{2}{\rm {Tr}} [{\cal U}(t)\rho(0){\cal U}(t)^{\dag}\lambda_i]
\label{r8}
\end{eqnarray} 
For the $\cal U$ matrix considered above, we have 
\begin{eqnarray}
\tau = \frac{2\pi n}{c}, \qquad n\in {\bf Z}
\label{r9}
\end{eqnarray} 
In the following we restrict ourselves to $n=1$. With this established, the $\cal F$ matrix can be evaluated and it takes the form 
\begin{eqnarray}
{\cal F}[{\cal U};\tau] = 
e^{\frac{-ia\,\tau}{2\sqrt{3}}} \left[
	\begin{array}{ccc}
		c\, \cos\frac{c\,\tau}{2}+i\sqrt{3}a\, \sin\frac{c\,\tau}{2} & 0 & 0 \\
		0 & e^{\frac{ia\sqrt{3}\,\tau}{2}} & 0\\
		 0 & 0&  e^{\frac{-ia\sqrt{3}\,\tau}{2}} \\
	\end{array}\right]
\label{r10}
\end{eqnarray}
with which the geometric phase follows by a direct evaluation of (30):
\begin{eqnarray}
\gamma_G [{\cal U}] = \arctan \left( \frac{\sin \left( \arctan \left[ \frac{1}{k}\tan \phi\right]\right)}{\frac{2\omega}{2\omega - 1}+ \cos (\arctan [k \tan \phi])}\right)
\label{r11}
\end{eqnarray}
where
\begin{eqnarray}
k= \frac{\sqrt{3} a}{c} \qquad {\rm and} \qquad \phi = \frac{\pi - 2c}{c\sqrt{3}}
\label{r12} 
\end{eqnarray}
In passing, we like to emphasize that a gauge transformation corresponding to any element of the little group ${\cal V} \in U(2)\times U(1)$ leaves the geometric phase invariant. For instance, if we had choosen 
\begin{eqnarray}
{\cal U}'(t) = {\cal U}(t) \cdot e^{-id\lambda_1t} \label{r13}
\end{eqnarray}   
which corresponds to a gauge transformed unitary matrix, then the $\cal F$ matrix takes the form 
\begin{eqnarray}  
{\cal F}[{\cal U'},\tau] = 
\frac{e^{\frac{-ia\tau}{2\sqrt{3}}}}{c}
\left[
\begin{array}{ccc}
		(a\sqrt{3}i \sin\frac{c\tau}{2}+c\cos\frac{c\tau}{2})\cos(d\tau)&(a\sqrt{3} \sin\frac{c\tau}{2}-ic\cos\frac{c\tau}{2})\sin(d\tau) & 0  \\
		-ice^{\frac{ia\tau\sqrt{3}}{2}}\,\sin(d\tau) &ce^{\frac{ia\tau\sqrt{3}}{2}}\,\cos(d\tau)  & 0\\
		0& 0&ce^{\frac{-ia\tau\sqrt{3}\,}{2}}\\		
	\end{array}\right]
\end{eqnarray}
Evaluation of the geometric phase yields expression (\ref{r11}), thereby demonstrating the gauge invariance of (30).  
\section{Concluding Remarks}
\par
To summarize briefly, we have considered the issue of phase holonomy of both non-degenerate and degenerate mixed states undergoing unitary evolution. Starting with the non-degenerate case, we have shown that the usual procedure of subtracting the dynamical phase from the total phase to yield the geometric phase,  does not hold for mixed states. To this end, we have furnished an expression for the geometric phase that is gauge invariant. In other words, unlike the procedure given in Ref.{\cite{Sjoqvist}}, we do not have to invoke the parallelity conditions to obtain the geometric phase from the total phase. In fact an obvious statement of gauge invariance is that any unitary matrix that realizes a path for the density matrix will yield a value that depends only on the path traversed in the state space. Moreover the form presented here also exhibits the parallelity conditions clearly. In essence, we have provided an explicit justification to the sufficient conditions furnished in Ref.{\cite{Sjoqvist}}. The analysis is also extended to cover states that may exhibit degeneracies. Here with the holonomy taking on a non-abelian character, we have provided an expression for the geometric phase that is manifestly gauge invariant. As in the case of the non-degenerate case, the form also displays the parallelity conditions which are shown to be generalizations of the ones provided in {\cite{Sjoqvist}}. We have also furnished explicit examples of the geometric phases for both the non-degenerate as well as degenerate mixed states. 
\centerline{\bf Acknowlegements}
We would like to thank Prof. C.H. Oh  whose guidance and valuable comments enabled the completion of this work. This work was supported in part by NUS Research Grant No. R-144-000-054-112 and A*STAR grant R-144-000-071-305.
\section{Appendix}
In this appendix we show that for a gauge transformation of the form (41), the expression for the geometric phase  (\ref{GT2}) remains unchanged. We proceed to prove this by establishing a few lemmas.
\vskip0.2cm
\noindent{\it{Lemma 1a}} 
\par\noindent
Let $X_1[{\cal U};\tau]$ and $X_2[{\cal U};\tau]$ be submatrices of $\rho_0{\cal U}(\tau)$ associated with the subspaces ${\cal H}_n$ and ${\cal H}_m$ respectively:
\begin{eqnarray*}
X_1[{\cal U};\tau]_{\mu\nu} &=& [\rho_0{\cal U}(\tau)]_{\mu\nu}= <\mu\vert \rho_0{\cal U}(\tau)\vert\nu>\qquad \mu,\nu = 1,2,...n \\
X_2[{\cal U};\tau]_{ij} &=& [\rho_0{\cal U}(\tau)]_{ij}= <i\vert \rho_0{\cal U}(\tau)\vert j>\qquad i,j = n+1,n+2,...N
\end{eqnarray*}
then 
$$
 {\rm Tr}_{{\cal H}}(\rho_0{\cal U}(\tau) {\cal F}[{\cal U;\tau}])=  {\rm Tr}_{{\cal H}_n}(X_1[{\cal U};\tau]{\cal F}_{{\cal H}_n}[{\cal U;\tau}])+ {\rm Tr}_{{\cal H}_m}(X_2[{\cal U};\tau]{\cal F}_{{\cal H}_m}[{\cal U;\tau}]).
$$
\vskip0.2cm
\noindent {\it Proof:} 
\par\noindent
The result follows immediately by taking
$${\rm Tr}_{{\cal H}}(\rho_0{\cal U}(\tau) {\cal F}[{\cal U;\tau}]) = \sum_{\mu=1}^{n}<\mu\vert \rho_0{\cal U}(\tau) {\cal F}[{\cal U;\tau}] \vert \mu> + \sum_{j=n+1}^{N}<j\vert \rho_0{\cal U}(\tau) {\cal F}[{\cal U;\tau}] \vert j>$$ 
and noting that ${\cal F}[{\cal U};\tau]$ is block diagonal (see (39)).
\vskip0.3cm
\noindent 
{\it Lemma 1b} 
\par\noindent
Under a gauge transformation (41), the submatrices $X_1[{\cal U};\tau]$ and $X_2[{\cal U};\tau]$ transform as follows:
\begin{eqnarray*}
X_1[{\cal U};\tau] \to X_1[{\cal U}';\tau] &=&  X_1[{\cal U};\tau]{\cal V}_1(\tau) \\
X_2[{\cal U};\tau] \to X_2[{\cal U}';\tau] &=&  X_2[{\cal U};\tau] \rm{diag}(e^{i\theta_{n+1}(\tau)},e^{i\theta_{n+2}(\tau)},...,e^{i\theta_N(\tau)})
\end{eqnarray*}
\noindent {\it Proof:} 
\par\noindent
This again follows from the block diagonal form of  ${\cal V}(t)$
\vskip0.3cm
\eject
\noindent 
{\it Lemma 2a} 
\par\noindent
Under a gauge transformation (41), ${\cal F}_{{\cal H}_m}[{\cal U};\tau]$ transforms as 
$$
{\cal F}_{{\cal H}_m}[{\cal U};\tau]  \to  {\cal F}_{{\cal H}_m}[{\cal U'};\tau]={\rm diag}(e^{i\theta_{n+1}(\tau)},e^{i\theta_{n+2}(\tau)},...,e^{i\theta_N(\tau)})^{\dag}{\cal F}_{{\cal H}_m}[{\cal U};\tau]
$$
\vskip0.2cm
\noindent {\it Proof:} 
\par\noindent
In proving this result, we will begin with (\ref{GT5b}). Now under ${\cal U}(t)\to {\cal U}'(t)= {\cal U}(t){\cal V}(t)$ we have
\begin{eqnarray*}
{\dot{\beta}'}_k(t)  & = &  i [{\cal U'}(t)^{\dag}\dot{{\cal U}}'(t)]_{kk} \\
& = & i[{\cal V}^{\dag}(t) {\cal U}(t)^{\dag}(\dot{\cal U}(t){\cal V}(t)+ {\cal U}(t)\dot{\cal V}(t))]_{kk}\\
& = & i[{\cal U}(t)^{\dag}\dot{{\cal U}}(t)]_{kk} - \dot{\theta}_k \qquad \quad k = n+1, n+2 ,... N.
\end{eqnarray*}
This in turn implies that 
\begin{eqnarray*}
 \frac{d}{dt}\left[{\beta}'_k + \theta_k \right] & = &  i[{\cal U}(t)^{\dag}\dot{{\cal U}}(t)]_{kk}  \\
& = & \frac{d}{dt}\left[{\beta}_k \right] 
\end{eqnarray*}
from which we surmise that 
$${\beta}'_k = {\beta}_k  - \theta_k + \xi_k $$
where $\xi_k$ is an arbitrary $t$-independent term. This term, however, vanishes when we impose initial conditions
$\beta_k(0)= \beta'_k(0)=0$ together with $\theta_k(0)=0$. 
Thus, we have for the $\cal F$- matrix
\begin{eqnarray*}
{\cal F}_{{\cal H}_m}[{\cal U}';\tau] & = & {\rm diag}(e^{i\beta'_{n+1}(\tau)},e^{i\beta'_{n+2}(\tau)},...,e^{i\beta'_N(\tau)})\\
& = & {\rm diag}(e^{i\theta_{n+1}(\tau)},e^{i\theta_{n+2}(\tau)},...,e^{i\theta_N(\tau)})^{\dag}{\cal F}_{{\cal H}_m}[{\cal U};\tau]
\end{eqnarray*}
which completes the proof.
\vskip0.3cm
\noindent 
{\it Lemma 2b} 
\par\noindent
Under a gauge transformation (41), ${\cal F}_{{\cal H}_n}[{\cal U};\tau]$ transforms as 
\begin{eqnarray*}
{\cal F}_{{\cal H}_n}[{\cal U};\tau] & \to & {\cal F}_{{\cal H}_n}[{\cal U'};\tau]= {\cal V}_1({\tau})^{\dag}{\cal F}_{{\cal H}_n}[{\cal U};\tau]
\end{eqnarray*}
\vskip0.2cm
\noindent {\it Proof:} 
\par\noindent
We first note that under ${\cal U}(t) \to {\cal U'}(t)= {\cal U}(t){\cal V}(t)$,  ${\cal A}\equiv {\cal U}(t)^{\dag}\dot{\cal U}(t)$ transforms as 
\begin{eqnarray*}
{\cal A} \to {\cal A'}  =  {\cal U'}(t)^{\dag}\dot{\cal U'}(t) = {\cal V}^{\dag} {\cal A} {\cal V} + {\cal V}^{\dag}\dot{\cal V}.
\end{eqnarray*}
Restricting ${\cal A}$ to the subspace ${\cal H}_n$, the $n\times n$ submatrix $\widetilde{\cal A}$ with components
\par\noindent
$\widetilde{\cal A}_{\mu\nu} = <\mu \vert{\cal A}\vert\nu >$ transforms as 
\begin{eqnarray*}
\widetilde{\cal A} \to \widetilde{\cal A}'  =   {\cal V}_1^{\dag} \widetilde{\cal A} {\cal V}_1 + {\cal V}_1^{\dag}\dot{\cal V}_1.
\end{eqnarray*}
 Then from (35), with ${\cal U} \to {\cal U}'$,
\begin{eqnarray*}
\dot{\alpha}' &=& - \widetilde{\cal A}'\alpha'\\
&=& - ({\cal V}_1^{\dag} \widetilde{\cal A} {\cal V}_1 + {\cal V}_1^{\dag}\dot{\cal V}_1
)\alpha'\\
&=& - {\cal V}_1^{\dag}(\widetilde{\cal A} {\cal V}_1 + \dot{\cal V}_1
)\alpha' 
\end{eqnarray*}
and thus 
\begin{eqnarray*}
\frac{d}{dt}\left[{\cal V}_1\alpha'\right] = -\widetilde{\cal A} {\cal V}_1 \alpha'.
\end{eqnarray*}
As a consequence of (35), we note that 
\begin{eqnarray*}
 {\cal V}_1 \alpha' = \alpha {\cal V}_0 \qquad \Longrightarrow \qquad \alpha'(t) = {\cal V}_1^{\dag}(t) \alpha (t){\cal V}_0
\end{eqnarray*}
where ${\cal V}_0$ is a $t$-independent but otherwise, arbitrary non-singular $n\times n$ matrix. As in the previous lemma, this arbitrariness is removed when we invoke the initial conditions $\alpha'(0)=\alpha(0)=1$, together with 
${\cal V}_1(0)=1$. As a result ${\cal V}_0 = 1$ and we have  
\begin{eqnarray*}
 {\cal F}_{{\cal H}_n}[{\cal U}';\tau] = \alpha'(\tau)= {\cal V}_1(\tau)^{\dag} \alpha(\tau) = {\cal V}_1(\tau)^{\dag}{\cal F}_{{\cal H}_n}[{\cal U};\tau]   
\end{eqnarray*}
which furnishes the desired result.
\vskip0.3cm
The gauge invariance of the geometric phase then follows immediately from the above lemmas:
\begin{eqnarray*}
 \gamma[{\cal U}'] &=& {\rm{arg}}\{ {\rm{Tr}}_{\cal H}(\rho_0 {\cal U}'(\tau) {\cal F}[{\cal U}';\tau])\}  \\
&=& {\rm{arg}}\{{\rm{Tr}}_{{\cal H}_n} (X_1[{\cal U}';\tau] {\cal F}_{{\cal H}_n}[{\cal U}';\tau])+ {\rm{Tr}}_{{\cal H}_m} (X_2[{\cal U}';\tau] {\cal F}_{{\cal H}_m}[{\cal U}';\tau]) \}\\
&=& {\rm{arg}}\{{\rm{Tr}}_{{\cal H}_n} (X_1[{\cal U};\tau] {\cal F}_{{\cal H}_n}[{\cal U};\tau])+ {\rm{Tr}}_{{\cal H}_m} (X_2[{\cal U};\tau] {\cal F}_{{\cal H}_m}[{\cal U};\tau]) \}\\
&=& {\rm{arg}}\{ {\rm{Tr}}_{\cal H}(\rho_0 {\cal U}(\tau) {\cal F}[{\cal U};\tau])\}\\
&=& \gamma[{\cal U}].
\end{eqnarray*}

\end{document}